\documentclass[twocolumn,10pt]{IEEEtran} 
\def\firstcompile{1}

\IEEEoverridecommandlockouts

\usepackage{cite}
\usepackage{amsmath,amssymb,amsfonts}
\usepackage{algorithmic}
\usepackage{graphicx}
\usepackage{textcomp}
\usepackage{xr}
\def\BibTeX{{\rm B\kern-.05em{\sc i\kern-.025em b}\kern-.08em
		T\kern-.1667em\lower.7ex\hbox{E}\kern-.125emX}}

\ifthenelse{\firstcompile=0}
{
	\usepackage{hyperref}
	\hypersetup{
		bookmarksdepth=2,
		bookmarks=true,     
		bookmarksopen=false, 
		hidelinks=true,
	}
}
{}

\usepackage{tikz}
\usetikzlibrary{backgrounds}
\usetikzlibrary{spy}
\usetikzlibrary{calc,arrows.meta}
\usepackage{pgfplots}
\usetikzlibrary{plotmarks}
\usetikzlibrary{arrows.meta}
\usetikzlibrary{positioning,fit}
\usetikzlibrary{intersections}
\usetikzlibrary{patterns}
\usetikzlibrary{decorations.shapes}

\usepgfplotslibrary{patchplots}
\usepgfplotslibrary{colormaps}

\usepackage{cancel}

\usepackage{pdfpages} 

\usepackage{multirow}
\usepackage{makecell}

\usepackage{algorithmic}
\usepackage[ruled,vlined]{algorithm2e}
\usepackage{setspace}
\SetKwRepeat{Do}{do}{while}
\SetKwRepeat{Until}{do}{until}

\pgfdeclarelayer{back1}
\pgfdeclarelayer{lay1}
\pgfdeclarelayer{back2}
\pgfdeclarelayer{lay2}
\pgfdeclarelayer{backback1}
\pgfsetlayers{backback1,back2,lay2,back1,lay1,main}

\ifthenelse{\firstcompile=2}
{
\newcommand{\exportFigures}{false}
\newcommand{\exportFiguresAsPNG}{false}
}
{
\newcommand{\exportFigures}{true}
\newcommand{\exportFiguresAsPNG}{true}
}

\ifthenelse{\equal{\exportFigures}{true}}
{
	\usepgfplotslibrary{external}
	\tikzexternalize[prefix=compiled_tikz_figures/,optimize command away=\includepdf]
	\ifthenelse{\equal{\exportFiguresAsPNG}{true}}
	{
		\tikzset
		{   png export/.style={
				external/system call={
					pdflatex \tikzexternalcheckshellescape -halt-on-error --extra-mem-top=10000000 -interaction=batchmode -jobname "\image" "\texsource" && pdftops -eps "\image.pdf" && convert -density 700 -transparent white "\image.pdf" "\image.png"
		}}}
		\tikzset{png export}
	}
	{}
}
{}

\usepackage{url}

\usepackage{bm}
\usepackage[caption=false,font=footnotesize]{subfig}
\usepackage[normalem]{ulem}

\usepackage{mathtools} 

\usepackage{acronym}

\usepackage{booktabs}

\usepackage[latin1]{inputenc}
\usepackage{tikz}
\usetikzlibrary{shapes,arrows}
\usetikzlibrary{arrows.meta}
\usetikzlibrary{positioning}

\usepackage{ellipsis}

\usetikzlibrary{calc}
\usetikzlibrary{decorations.pathreplacing,decorations.markings,shapes.geometric}
\usetikzlibrary{decorations.pathmorphing}
\usetikzlibrary{fit}
\usetikzlibrary{pgfplots.groupplots}

\usetikzlibrary{calc,arrows.meta}

\usetikzlibrary{backgrounds} 

\definecolor{green(pigment)}{rgb}{0.0, 0.65, 0.31}
\definecolor{frenchblue}{rgb}{0.0, 0.45, 0.73} 
\definecolor{mediumcandyapplered}{rgb}{0.89, 0.02, 0.17}

\usepackage[most]{tcolorbox}
\tcbuselibrary{breakable}
\tcbset{every box/.style={enhanced,breakable}}
\tcbset{colframe=black,colback=red!10,enhanced,breakable,sharp corners}

\usepackage{enumitem} 

\usepackage{notation}

\usepackage{adjustbox}

\definecolor{alex}{RGB}{51,183,150}
\definecolor{erik}{RGB}{235,134,52}

\newcommand{\ticked}{$\text{\rlap{$\checkmark$}}\square$}
\newcommand{\unticked}{{$\square$}}
\newcommand{\tick}[1]{\ifthenelse{#1=1}{\ticked}{\unticked}}

\graphicspath{{./figures/}}

\allowdisplaybreaks

\newcommand{\rmv}{\hspace*{-.3mm}}

\renewcommand{\exp}[1]{\ensuremath{{e}^{#1}}}

\newcommand{\s}{\hspace*{0.5pt}}

\newcommand{\ist}{\hspace*{.3mm}}
\newcommand{\iist}{\hspace*{1mm}}

\newcommand{\nn}{\nonumber}

\newcommand{\atantwo}{\text{atan2}}

\newcommand{\zd}{\ensuremath{z^{(i)}_{\text{d}_{m,n}}}}
\newcommand{\zu}{\ensuremath{z^{(i)}_{\mathrm{u}_{m,n}}}}

\newcommand{\zaoa}{\ensuremath{z^{(i)}_{\theta_{m,n}}}}
\newcommand{\zaod}{\ensuremath{z^{(i)}_{\vartheta_{m,n}}}}

\newcommand{\T}{\text{T}}

\DeclareMathOperator*{\multisum}{\sum \cdots \sum}

\begin{document}

\title{\huge{Simultaneous Source Separation, Synchronization, Localization and Mapping for 6G Systems\\[-1mm]}}	
\author{\normalsize \IEEEauthorblockN{Alexander Venus,~\IEEEmembership{\normalsize  Member,~IEEE}, Erik Leitinger,~\IEEEmembership{\normalsize Member,~IEEE}, and Klaus Witrisal,~\IEEEmembership{\normalsize Member,~IEEE}\\[0mm]}
	\IEEEauthorblockA{
		Institute of Comm. Networks and Satellite Comms., Graz University of Technology, Austria \\ 
	}
	\vspace*{-8mm}}

\maketitle
\frenchspacing

\renewcommand{\baselinestretch}{0.96}\small\normalsize

\begin{abstract}
	\Ac{mpslam} is a promising approach for future 6G networks to jointly estimate the positions of transmitters and receivers together with the propagation environment. In cooperative \ac{mpslam}, information collected by multiple \acp{mt} is fused to enhance accuracy and robustness. 
Existing methods, however, typically assume perfectly synchronized \acp{bs} and orthogonal transmission sequences, rendering inter-\ac{bs} interference at the \acp{mt} negligible. 
In this work, we relax these assumptions and address simultaneous source separation, synchronization, and mapping. A relevant example arises in modern 5G systems, where \acp{bs} employ muting patterns to mitigate interference, yet localization performance still degrades. 
We propose a novel \ac{bs}-dependent data association and synchronization bias model, integrated into a joint Bayesian framework and inferred via the sum-product algorithm on a factor graph. 
The impact of joint synchronization and source separation is analyzed under various system configurations. Compared with state-of-the-art cooperative \ac{mpslam} assuming orthogonal and synchronized \acp{bs}, our statistical analysis shows no significant performance degradation. 
The proposed BS-dependent data association model constitutes a principled approach for classifying features by arbitrary properties that persist over time, such as reflection order or feature type (scatter points versus walls).
\thispagestyle{empty} 
\end{abstract}

\acresetall 
\IEEEpeerreviewmaketitle

\vspace{-6.5mm} 
\section{Introduction}\label{sec:introduction}
\begin{figure}[t]
	\centering
	\scalebox{0.9}{\includegraphics{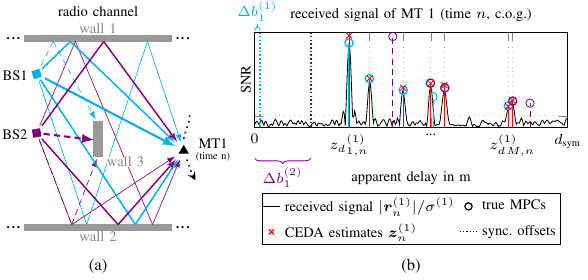}}
	\vspace{-8mm}
	\caption{Figure (a) shows an exemplary environment consisting of three walls including a single \ac{mt} that receives radio signals via multipath propagation from two \textit{unsynchronized, interfering} \ac{bs} that transmit using identical frequency bands/symbols. Figure (b) shows the received signal at the center of gravity of MT1 at time $n$. The signals from BS1 and BS2 are delayed with respect to time~$0$ by the apparent synchronization biases $\Delta b_{1}^{(1)} = b_{\mathrm{bs},n}^{(1)} - b_{\mathrm{mt},n}^{(1)}$ and $\Delta b_{1}^{(2)} = b_{\mathrm{bs},n}^{(2)} - b_{\mathrm{mt},n}^{(1)}$, which arise from the differences between the corresponding BS and MT clock biases.
	}\label{fig:eye_catcher}
	\vspace{-6mm}
\end{figure}

Location awareness is a key ingredient for future 6G communication networks, enabling adaptive beamforming, efficient resource allocation, and interference management in the context of integrated communication and sensing (ISAC) \cite{Wymeersch2024,GonFurKalValDarSheSheBayWymProcIEEE2024}.  
Yet, radio-based localization of \acp{mt} remains challenging in indoor and urban environments \cite{WitMeiLeiSheGusTufHanDarMolConWin:J16, MenWymBauAbu:TWC2019}, which are dominated by strong multipath propagation and frequent \acl{olos} conditions (see Fig.~\ref{fig:eye_catcher}). These effects can lead to localization outages and, consequently, degrade communication performance in ISAC systems. 

In contrast to conventional methods that treat multipath as interference, \ac{mpslam} exploits it as an additional source of information by modeling specular reflections as \acfp{va}---the mirror images of \acp{bs} at reflecting surfaces \cite{WitMeiLeiSheGusTufHanDarMolConWin:J16}. The joint estimation of \acp{mt} and \acp{va} thus enables accurate and robust localization even in \acl{olos} conditions \cite{GentnerTWC2016, LeiMeyHlaWitTufWin:J19 , KimGarGeXiaSveWym:TSP2024,VenLeiTerWit:TWC2023,VenLeiTerMeyWit:TWC2024,LiaLeiMey:TSP2025,GeKalXiaGarAngKimTalValWymSev:TSP2025}.

\paragraph*{State of the Art}

\Ac{mpslam} is a feature-based \acs{slam} method \cite{DurrantWhyte2006}, where the environment is represented by static map features---namely, \acp{bs} and \acp{va}. Since both the number and positions of these features are unknown, \ac{mpslam} jointly estimates them together with the agent trajectory.  
The input measurements (such as \ac{toa}, \ac{aoa}, \ac{aod} of \acp{mpc}) are typically obtained from radio signals via parametric channel estimation algorithms \cite{HanFleRao:TSP2018, GreLeiWitFle:TWC2024, MoeWesVenLei:Fusion2025}.  
The probabilistic formulation of \ac{mpslam} is commonly derived through belief propagation on a factor-graph representation of the SLAM problem \cite{LeiMeyHlaWitTufWin:J19}, leading to scalable algorithms suitable for various network scenarios.  
It has been shown that \ac{mpslam} constitutes an instance of marginal Poisson multi-Bernoulli (PMB) SLAM, offering an excellent trade-off between estimation robustness and computational scalability compared to other random finite set (RFS)-based SLAM approaches \cite{KimGraSveKimWym:TVT2022}.  
Recent extensions include cooperative simultaneous localization and tracking (SLAT) \cite{Brambilla2022} and cooperative SLAM within the \ac{mpslam} framework \cite{LeiWieVenWit:Asilomar2024}.

When multiple \acp{bs} are involved, existing \ac{mpslam} approaches typically assume perfect clock synchronization among the \acp{bs} and orthogonal transmission sequences, such that inter-\ac{bs} interference at the \ac{mt} is negligible.  
A particularly relevant example arises in modern 5G systems, where positioning reference signals (PRS) are used for downlink localization. A major challenge in this context stems from unsynchronized \acp{bs}, which transmit signals with irregular timing, thereby causing interference at the \acp{mt} \cite{HuBerLiRus:Globecom2017} (see Fig.~\ref{fig:eye_catcher}).  
Although the PRS standard employs muting patterns to mitigate interference, localization performance and robustness still degrade significantly under such conditions.

\begin{figure*}[t]
	\centering
	\scalebox{0.9}{\includegraphics{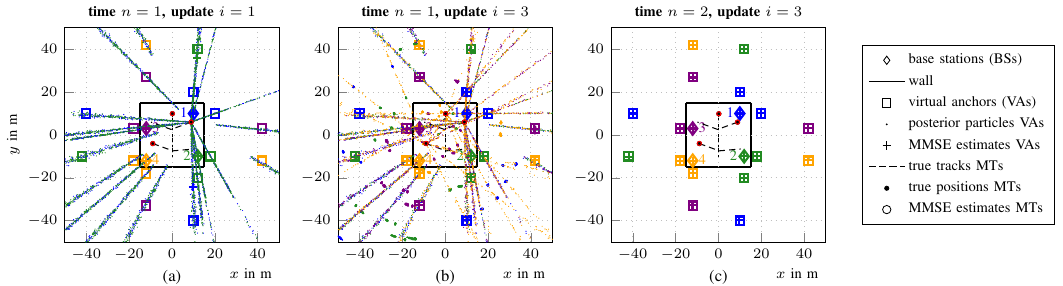}}
	\vspace{-2mm}
	\caption{Simulation environment for time $n=1$ after the first MT update (a), the last MT update (b) and for time $n=2$ after the last MT update. The colors represent the true (box) and the estimated (+ and dot) association of map features (\acp{pva}) to \acp{bs}.}\label{fig:converging_pvas}
	\vspace{-6mm}
\end{figure*}

\paragraph*{Contributions}

In this work, we present a robust and scalable algorithm for \emph{simultaneous source separation, synchronization, and mapping}. The proposed \ac{mpslam} algorithm processes downlink radio signals transmitted from multiple unsynchronized \acp{bs} to multiple unsynchronized \acp{mt}, enabling sequential estimation of multiple moving \acp{mt} while jointly detecting and estimating the locations of map features---i.e., \acp{bs} and \acp{va}---represented by their \acp{pva} positions and existence probabilities. Since each \ac{mt} receives a mixture of potentially interfering impulse responses from all visible \acp{bs} (see Fig.~\ref{fig:eye_catcher}), the algorithm performs probabilistic source separation by associating \acp{pva} to their corresponding \ac{bs} (cf. Fig.~\ref{fig:converging_pvas}). It further conducts \ac{pda} between \acp{mpc} measurements and \acp{pva}, while jointly estimating \ac{bs}-related and \ac{mt}-related synchronization offsets. This enables consistent data fusion of map features across all unsynchronized and mutually interfering BS--MT communication links. The key contributions are summarized as follows:
\begin{itemize}
	\item Introduction of a novel data association model for interfering \acp{bs}, enabling probabilistic source separation.  
	\item Modeling of apparent synchronization bias as a difference between \ac{bs}- and \ac{mt}-related random variables.  
	\item Derivation of the joint posterior distribution and corresponding factor-graph-based \ac{spa} for the complete problem.  
	\item Performance analysis demonstrating the impact of joint synchronization and source separation, and comparison with state-of-the-art cooperative \ac{mpslam} approaches.  
\end{itemize}

\vspace{-2mm} 
\section{Geometrical Relations}\label{sec:geometric_model}
\begin{figure}[t]
	\centering   
	\scalebox{0.8}{\includegraphics{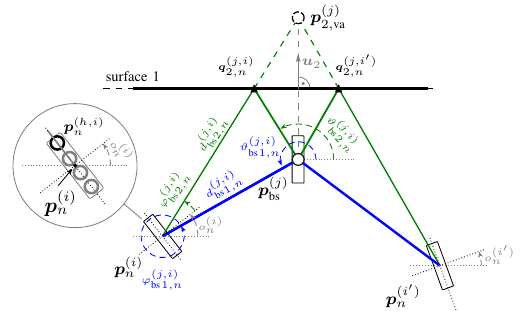}}
	\vspace*{-3mm}
	\caption{Exemplary indoor environment with one \ac{bs} at position \smash{$\V{p}^{(j)}_\text{bs} $} and two MTs at position $\V{p}_{n}^{(i)}$ and $\V{p}_{n}^{(i')}\rmv\rmv$, jointly observing a \ac{va} at $\V{p}^{(j)}_{\text{va},l}$. The figure also illustrates the array geometry of the agents and \acp{pa}, assuming $o^{(j)}_{n}=0$. }
	\vspace*{-4mm}
	\label{fig:geometric_relations}
\end{figure}

At time step $n$, we consider $I$ \acp{mt} with unknown state $\V{x}^{(i)}_{n}$ containing the positions $\V{p}^{(i)}_{n} = [p^{(i)}_{x,n} \iist p^{(i)}_{y,n}]^\text{T}$, the velocity $\V{v}^{(i)}_{n} = [v^{(i)}_{x,n} \iist v^{(i)}_{y,n}]^\text{T}$, and antenna array orientation $o^{(i)}_{n}$ where $i \in \{1, \dots,I\}$. We consider $J$ \acp{bs} with known positions $\V{p}_{\mathrm{bs}}^{(j)}, j \in \{1, \dots, J\}$. Each \ac{bs} $j$ observes $N_{n}^{(j)} - 1$ specular reflections of radio signals at flat surfaces, modeled by \acp{va} with unknown positions $\V{p}_{\mathrm{va},l}^{(j)} = \big[ {p}_{x,\mathrm{va},l}^{(j)} \ist\ist\ist {p}_{y,\mathrm{va},l}^{(j)} \big]^{\mathrm{T}} , l \in \{2, \dots, N_{n}^{(j)}\}$ (see \cite{LeiVenTeaMey:TSP2023}). By applying the image-source model \cite{Bor:JASA1984,LeiVenTeaMey:TSP2023}, a \ac{va} associated with a single-bounce path is the mirror image of $\V{p}_{\mathrm{bs}}^{(j)}$ at reflective surface $l$ given by $\V{p}^{(j)}_{\mathrm{va},l} = \V{p}^{(j)}_{\mathrm{bs}} + 2\big( \V{u}_l^{\T}\V{e}_l - \V{u}_l^{\T}\V{p}^{(j)}_{\mathrm{bs}}\big)\V{u}_l$
where $\V{u}_l$ is the normal vector of reflective surface $l$, and $\V{e}_l$ is an arbitrary point on the considered surface.\footnote{We consider only \acp{va} associated to single-bounce reflections; extensions to double-bounce reflections can be done in accordance with \cite{LeiVenTeaMey:TSP2023,LiCaiLeiTuf:ICC2024}.}
The according point of reflection $\V{q}_{l,n}^{(j,i)}$ at the surface is given as
\vspace*{-2mm}
\begin{align}
	\V{q}_{l,n}^{(j,i)} = \V{p}_{\text{va},l}^{(j)} + \frac{(\V{p}_{\text{bs}}^{(j)} - \V{p}_{\text{va},l}^{(j)})^\T \V{u}_l}{2(\V{p}^{(i)}_{n} - \V{p}_{\text{va},l}^{(j)})^\T \V{u}_l} (\V{p}^{(i)}_{n} - \V{p}_{\text{va},l}^{(j)})
	\label{eq:reflectionPoint}\\[-7mm]\nn
\end{align}
and is needed to relate \ac{aod} of a specular reflection to the corresponding \ac{va}. Both the \acp{mt} and all \acp{bs} are equipped with antenna arrays. As shown in Fig.~\ref{fig:geometric_relations}, the antenna array geometry at the \ac{bs} is defined by array element positions relative to \(\V{p}_{\mathrm{bs}}^{(j)}\) (with known orientation of zero), and for an \ac{mt} by positions relative to \(\V{p}^{(i)}_{n}\) with unknown orientation \(o^{(i)}_{n}\). 
The geometric relationships of the \ac{mpc} parameters -- distance, \ac{aoa}, and \ac{aod} of \ac{va} $k$ for \ac{mt} $i$ and \ac{bs} $j$ and are given by 
$d_{{l,n}}^{(j,i)} \triangleq \|\V{p}^{(i)}_{n} \!-\rmv \V{p}_{\text{va},l}^{(j)}\|$,  
$\varphi^{(j,i)}_{{l,n}} \triangleq \angle(\V{p}^{(i)}_{n},\V{p}_{\text{va},l}^{(j)}) - o^{(i)}_{n}$ 
with $ \angle(\V{p}_{1},\V{p}_{2}) = \atantwo\big(p_{y,2}-p_{y,1},p_{x,2}-p_{x,1})$ and $\vartheta^{(j,i)}_{{l,n}} \triangleq \angle(\V{q}_{l,n}^{(j,i)},\V{p}^{(i)}_{n})$. The according geometric relationships of the \ac{los} parameters are given by $d_{\text{bs},n}^{(j,i)} \triangleq \|\V{p}^{(i)}_{n} \!-\rmv \V{p}_{\text{bs}}^{(j)}\|$, $\varphi^{(j,i)}_{\text{bs},n} \triangleq \angle(\V{p}^{(i)}_{n},\V{p}_{\text{bs}}^{(j)}) - o^{(i)}_{n}$ and $\vartheta^{(j,i)}_{\text{bs},n} \triangleq \angle(\V{p}_{\text{bs}}^{(j)},\V{p}^{(i)}_{n}) -  o^{(j)}_{\text{bs}}$, where $o^{(j)}_{\text{bs}}$ is known (see Fig.~\ref{fig:geometric_relations}).

\vspace{-1mm} 
\section{Radio Signal Model and Channel Estimation}\label{sec:signal_model}
Each \ac{bs} $j$ continuously transmits an \ac{rf} pilot signal $s(t)$ of bandwidth $B$ at center frequency $f_\mathrm{c}$. Each \ac{mt} $i$ acts as a receiver, observing both, the \ac{los} component and \acp{mpc} caused by reflections from surrounding objects. Since all \acp{bs} employ the same pilot sequence, every \ac{mt} observes an inseparable superposition of signals from all \acp{bs}.

We consider a MIMO system in which each \ac{mt}~$i$ and each \ac{bs}~$j$ is equipped with an antenna array comprising $H$ and $H'$ elements, respectively. The array elements are located at positions $\V{p}^{(h,i)}_{n} \in \mathbb{R}^2$ for the \ac{mt} (or $\V{p}^{(h,j)} \in \mathbb{R}^2$ for the \ac{bs}), with $h \in \{1,\ldots,H\}$ (or $h \in \{1,\ldots,H'\}$), as illustrated in Fig.~\ref{fig:geometric_relations}. The array center of gravity is given by $\V{p}^{(i)}_{n} = \frac{1}{H}\sum_{h=1}^{H} \V{p}^{(h,i)}_{n}$ for the \ac{mt} (or $\V{p}^{(j)}_{\text{bs}}$ for the \ac{bs}). The array orientation of \ac{mt}~$i$ at time~$n$ is denoted by $o^{(i)}_{n}$, while, without loss of generality, the orientation of each \ac{bs} is fixed to $o^{(j)}_{\text{bs}} = 0$. 
We further assume that \acp{bs} and \acp{mt} are mutually unsynchronized. In particular, \ac{bs} $j$ has a clock bias $\vspace*{-0.6mm}b_{\mathrm{bs},n}^{(j)}$ and \ac{mt} $i$ has a clock bias $ b_{\mathrm{mt},n}^{(i)}\vspace*{-0.6mm}$. At each discrete time index $n$, the $I$ \acp{mt} sample the received waveform synchronously at rate $B$. In frequency domain, this results in $M = B/\Delta$ samples with frequency spacing $\Delta$. The discrete-frequency \ac{rf} signal model between \ac{bs} $j$ and \ac{mt} $i$ is given by \cite{GreLeiWitFle:TWC2024}
\vspace*{-2mm}
\begin{align}\label{eq:signal_modelBSMT}
	{r}_{n}^{(i)}[\ell,h,h'] &= \sum_{j=1}^{J} \sum_{l=1}^{{N}_n^{(j,i)}} {\alpha}_{l,n}^{(j,i)} \exp{\big(i 2 \pi \frac{f_\mathrm{c}}{c} d^{(j,h)} 
		\cos({\vartheta}^{(j,i)}_{l,n} -\psi^{(j,h)})\big)} \nn \\[-1mm]
	& \hspace{2mm}\times\exp{\big(i 2 \pi \frac{f_\mathrm{c}}{c} d_n^{(i,h')} 
		\cos({\varphi}^{(j,i)}_{l,n}- o_{i,n} -\psi_n^{(i,h')})\big)} \nn \\[-1mm]
	&  \hspace{2mm} \times {s}({\tau}_{l,n}^{(j,i)})[\ell] + \ist {w}_n^{(j,i)}[\ell,h,h']
\end{align}
where  ${s}(\tau)[\ell] =  S^{(j)}(\ell \Delta) \exp{\big(i 2\pi \ell \Delta (\tau + {b_\text{bs}}_{n}^{(j)} - {b_\text{mt}}_{n}^{(i)}) \big)} $ with ${b_\text{bs}}_{n}^{(j)}~-~{b_\text{mt}}_{n}^{(i)}$ being the observed clock bias. ${{N_\text{bs}}_n^{(j,i)}}$ is the number of visible path (related to \acp{va}), $S^{(j)}(f)$ represents the signal spectrum, which is sampled at $\ell =-(M-1)/2,\iist\ldots\iist,(M-1)/2$, $h$ represent the index for \ac{bs} antennas\footnote{
	In 5G systems, access to the impulse responses of individual antenna elements is typically unavailable. Instead, beamforming is performed at the \ac{bs}, and spreading sequences of individual beams are embedded within dedicated resource blocks (RBs). The signal model in \eqref{eq:signal_modelBSMT} can straightforwardly be reformulated in the beam space following \cite{Xu1994}, allowing for equivalent application of parametric channel estimation according to Section~\ref{sec:channel_estimation}. Note that we assume the spreading sequences corresponding to different beams to be encoded in RBs within the same PRS symbol. If this assumption is dropped, data association must be carried out individually for each beam, since the inter-BS interference pattern then differs across beams. 
}, $h'$ represents the index for \ac{mt} antennas, and $c$ is the speed-of-light. The quantities $d_n^{(i,h')}$ and $\psi_n^{(i,h')}$ (respectively,
$d^{(j,h')}$ and $\psi^{(j,h')}$) denote the propagation delay and angle associated with the individual antenna element located at position
$\mathbf{p}_n^{(h,i)}$ (respectively, $\mathbf{p}^{(h,j)}$) relative to the center of gravity of the corresponding array.
${\tau}_{l,n}^{(j,i)}$, $\varphi^{(j,i)}_{l,n}$ and $\vartheta^{(j,i)}_{l,n}$ denote respectively delay, \ac{aoa} and \ac{aod} of the \acp{mpc}. The complex amplitude $\alpha_{l,n}^{(j,i)} \in \mathbb{C}$ is given by ${\alpha}_{l,n}^{(j,i)} = {a}^{(j,i)}_{l,n}\ist \text{e}^{-j2\pi f_\text{c}  {\tau}_{l,n}^{(j,i)}} c/(4\pi f_\text{c} {d}_{l,n}^{(j,i)})$ where ${a}^{(j,i)}_{l,n} \in \mathbb{C}$ is a reflection coefficient originating from all interactions of the radio signal with the radio equipment and the associated flat surfaces. The last term ${w}_n^{(j,i)}[\ell,h]$  aggregates the measurement noise. The according received signal vector is $\V{r}^{(i)}_{n} = \big[\V{r}_{n}^{(i)\ist\text{T}}[1,1] \hspace{1mm} \cdots \hspace{1mm} \V{r}_{n}^{(i)\ist\text{T}}[H,1] \iist \V{r}_{n}^{(i)\ist\text{T}}[1,2] \hspace{1mm} \cdots \hspace{1mm} $ $\V{r}_{n}^{(i)\ist\text{T}}[H,H']\big]^\T\in \mathbb{C}^{MHH'}$.

\vspace{-2mm} 
\subsection{Parametric Channel Estimation: Measurements}  \label{sec:channel_estimation}

At each time step $n$ for each \ac{mt} $i$, we apply a \ac{ceda} \cite{HanFleRao:TSP2018,GreLeiWitFle:TWC2024,MoeWesVenLei:Fusion2025} to the discrete MIMO signal vectors, yielding \ac{mpc} parameters estimates $\V{z}^{(i)}_{m,n} = [\zd \;\zaoa \;\zaod \;\zu ]^{\mathrm{T}}$, where $m\in\mathcal{M}^{(i)}_{n}\triangleq\{1,\dots,M^{(i)}_{n}\}$. Each $\V{z}^{(i)}_{m,n}$ contains $\zd \in[0,\tau_\text{max}]$, $\zaoa \in[-\pi,\pi]$, $\zaod\in[-\pi,\pi]$, and $\zu\in[\gamma,\infty)$, where $\gamma$ is the detection threshold. The stacked measurement vector $ \bm{z}^{(i)}_{n} = [{\bm{z}^{(i)\text{T}}_{1,n}}  ... \, {\bm{z}^{(j)\text{T}}_{M_n^{(i)},n}}]^\text{T}$ collects all estimates -- each being potentially caused by any interfering \acp{bs} $j \in \{1,\dots,J\}$ (or due to a \ac{fp} of the \ac{ceda}).

\vspace{-1mm} 
\section{System Model} \label{sec:system_model}
At each time $n$, the proposed method infers random variables explicitly denoted in san serif, upright font. Their definitions are in accordance with Sections~\ref{sec:geometric_model} and \ref{sec:signal_model}. The states of the \acp{mt} are given by $\RV{x}_{n}^{(i)} = [\tilde{\RV{x}}^{(i)\ist\T}_{n} \ist \RV{v}_{n}^{(i)\ist\T}~\rv{b}_{\text{mt}\, n}^{(i)}]^\text{T}$, with the sub-state 
$\tilde{\RV{x}}^{(i)\ist\T}_{n} = [\RV{p}_{n}^{(i)\ist\T} \ist\rv{o}_{n}^{(i)}]^\text{T}$. 
The BS biases $\rv{b}_{\text{bs},n}^{(j)}$ are collected by $\RV{b}_{\text{bs},n}^{(j)} = [\rv{b}_{\text{bs}, n}^{(1)} \iist \cdots \iist \rv{b}_{\text{bs}, n}^{(J)}]^\text{T}$.

As in \cite{MeyKroWilLauHlaBraWin:J18, LeiMeyHlaWitTufWin:J19, LeiVenTeaMey:TSP2023}, we account for the unknown number of \acp{va} by introducing \acfp{pva} $k \in \mathcal{K}_n \triangleq \{1,\dots,\rv{K}_n\}$. The number $\rv{K}_n$ of \acp{pva} of \textit{all interfering} \acp{bs} $j \in \{1 ,\dots, J\}$ is the maximum possible number of actual \acp{va}, i.e., all \acp{va} that produced a measurement so far \cite{MeyKroWilLauHlaBraWin:J18, LeiMeyHlaWitTufWin:J19, LeiVenTeaMey:TSP2023} (where $\rv{K}_n$ increases with time $n$). The states of the \acp{pva} are given by $\RV{y}_{k,n} \rmv =\rmv \big[\RV{p}^{\T}_{\text{va},k} \; \rv{r}_{k,n} \big ]^\T$. 
The existence/nonexistence of \acp{pva} $k$ is modeled by the existence variable $r_{k,n} \in \{0,1\}$ in the sense that \acp{pva} exists if $r_{k,n} = 1$. Formally, even if \acp{pva} $k$ is nonexistent, i.e., if $r_{k,n} = 0$, its state is considered. 
The states $\RV{p}_{\text{va},k}$ of nonexistent \acp{pva} are obviously irrelevant. Therefore, all \acp{pdf} defined for \ac{pva} states, $f(\V{y}_{k,n}) \rmv\rmv=\rmv\rmv f(\V{p}_{\text{va},k}, r_{k,n})$, are of the form $f(\V{p}_{\text{va},k}, 0 )\rmv\rmv=\rmv\rmv f_{k,n} f_{\text{d}}(\V{p}_{\text{va},k})$, where $f_{\text{d}}(\V{p}_{\text{va},k})$ is an arbitrary ``dummy \ac{pdf}'' and $f_{k,n} \!\rmv\in [0,1]$ is a constant and can be interpreted as the probability of non-existence \cite{MeyKroWilLauHlaBraWin:J18, LeiMeyHlaWitTufWin:J19}. 
%


\vspace{-3.5mm} 
\subsection{Measurement Model and New PVAs}\label{sec:measmodslam3}
\vspace{-1.5mm}
An existing \ac{pva} generates a measurement $\RV{z}^{(i)}_{m,n}\rmv\rmv\rmv\rmv\in  \{1,\dots,$\linebreak $\rv{M}^{(i)}_{n}\}$ with detection probability $p_{\mathrm{d}}$. 
The single-measurement \ac{lhf} $f({\V{z}_\text{bs}}^{(i)}_{m,n}|\V{x}_{n}^{(i)},\V{p}_{\text{va},k},{b}^{(j)}_{\text{bs},n})$ is assumed conditionally independent across the entries of ${\V{z}_\text{bs}}^{(i)}_{m,n}$ and factorizes as
\vspace*{-2mm}
\begin{align}
	&f({\V{z}}^{(i)}_{m,n}|\V{x}_{n}^{(i)}, \V{p}_{\text{va},k}\rmv, {b}^{(j)}_{\text{bs},n})= f({z_\mathrm{d}}^{(i)}_{m,n} |\V{p}_{n}^{(i)}, \V{p}_{\text{va},k}\rmv, {b}^{(j)}_{\text{bs},n})  \nn\\
	&\hspace*{15mm}\times f({z_\mathrm{\varphi}}^{(i)}_{m,n} |\tilde{\V{x}}_n^{(i)}, \V{p}_{\text{va},k}) 
	f({z_\mathrm{\vartheta}}^{(i)}_{m,n} |{\V{p}}_n^{(i)},\V{p}_{\text{va},k}) \\[-7mm]\nn
\end{align}
The \acp{lhf} of distance, \ac{aoa}, and \ac{aod} are Gaussian, $f_{\mathrm{N}}(x;\mu,\sigma^2)$, with
\vspace*{-2mm}
\begin{align}
	f({z_\mathrm{d}}^{(i)}_{m,n} | \V{p}_{n}^{(i)}, \V{p}_{\text{va},k}, {b}^{(j)}_{\text{bs},n}) &= f_{\mathrm{N}}({z_\mathrm{d}}^{(i)}_{m,n}; d^{(k,i)}_{k,n}  \nn\\
	&\hspace*{7mm} - ( {b}_{\text{mt}\, n}^{(i)} -  {b}_{\text{bs}\, n}^{(j)} ), \sigma_{\mathrm{d}\, k,n}^{(i)\s 2} )
	\label{eq:LHF_dist3}\\
	f({z_\mathrm{\varphi}}^{(i)}_{m,n}| \tilde{\V{x}}_{n}^{(i)}, \V{p}_{\text{va},k}) &= f_{\mathrm{N}}({z_\mathrm{\varphi}}^{(i)}_{m,n}; \varphi^{(k,i)}_{k,n} , \sigma^{(k,i)\ist2}_{\varphi_{k,n}}) 
	\label{eq:LHF_aAoA} \\
	f({z_\mathrm{\vartheta}}^{(i)}_{m,n} | {\V{p}}_n^{(i)}, \V{p}_{\text{va},k}) &= f_{\mathrm{N}}({z_\mathrm{\vartheta}}^{(i)}_{m,n};\vartheta^{(k,i)}_{k,n}, \sigma^{(k,i)\ist2}_{\vartheta_{k,n}} ) 
	\label{eq:LHF_aAoD}\\[-6.5mm]\nn
\end{align}
where the means $d^{(k,i)}_{k,n}$, $\varphi^{(k,i)}_{k,n}$ and $\vartheta^{(k,i)}_{k,n}$ are defined in Section~\ref{sec:geometric_model}, and the variances, derived from the Fisher information, are 
$\sigma^{(i)\,2}_{d_{k,n}} = c^2/(8\pi^2\beta_{\mathrm{bw}}^2{z_\mathrm{u}}^{(i)\,2}_{m,n})$, 
${\sigma_{\varphi}}^{(k,i)\,2}_{k,n} = 1/\!\big(8\pi^2{z_\mathrm{u}}^{(i)\,2}_{m,n}D^2(\varphi^{(k,i)}_{k,n})\big)$, and 
${\sigma_{\vartheta}}^{(k,i)\,2}_{k,n} = 1/\!\big(8\pi^2{z_\mathrm{u}}^{(i)\,2}_{m,n}D^2(\vartheta^{(k,i)}_{l,n})\big)$.
Here, $\beta_{\mathrm{bw}}^2$ denotes the mean square bandwidth and $D^2(\cdot)$ the normalized squared array aperture \cite{FacDeuKesWilColWitLeiSecWym:ICC2023,LiCaiLeiTuf:ICC2024}.
Similarly, for the \ac{bs}-related \ac{lhf}, we get
\vspace*{-1mm}
\begin{align}
	&f_{\text{bs}}({\V{z}}^{(i)}_{m,n}|\V{x}_{n}^{(i)}, {b}^{(j)}_{\text{bs},n})= f({z_\mathrm{d}}^{(i)}_{m,n} |\V{p}_{n}^{(i)}, {b}^{(j)}_{\text{bs},n})  \nn\\
	&\hspace*{30mm}\times f({z_\mathrm{\varphi}}^{(i)}_{m,n} |\tilde{\V{x}}_n^{(i)}) 
	f({z_\mathrm{\vartheta}}^{(i)}_{m,n} |\V{p}_{n}^{(i)}) \\[-6.5mm]\nn
\end{align}
where within the indivual \acp{lhf} the geometrical relations $d^{(k,i)}_{k,n}$, $\varphi^{(k,i)}_{k,n}$, and $\vartheta^{(k,i)}_{k,n}$ are replaced by $d_{\text{bs},n}^{(j,i)}$, $\varphi^{(j,i)}_{\text{bs},n}$, and $\vartheta^{(j,i)}_{\text{bs},n}$, respectively.

Detected \acp{va}, i.e., \acp{va} that generated a measurement for the first time, are modeled by a Poisson process with mean $\mu_{\mathrm{n}}$ and \ac{pdf} $f_{\mathrm{n}}( \overline{\V{p}}^{(j,i)}_{\text{va},m}|\V{x}_{n}^{(i)})$. 
Accordingly, newly detected \acp{va} are represented by \textit{new \ac{pva} states} $\overline{\RV{y}}^{(j,i)}_{n,m}=[\overline{\RV{p}}^{(j,i)\ist\T}_{\text{va},m}  \iist \overline{\rv{r}}^{(j,i)}_{m}]^{\T}$, $m \rmv\rmv \in \{1,\dots,  \rv{M}^{(i)}_{n} \}$ , $j  \in \{1,\dots,  J\}$ in our statistical model. Each new \ac{pva} state corresponds to a measurement ${\RV{z}_\text{bs}}^{(i)}_{m,n}$; $\overline{r}_{m,n} \!=\! 1$ implies that measurement ${\RV{z}_\text{bs}}^{(i)}_{m,n}$ was generated by a newly detected \ac{va}. 
\textit{It is also unknown which \ac{bs} (if any) caused a measurement ${\RV{z}_\text{bs}}^{(i)}_{m,n}$. Thus, for each index $m$, a number of $J$ new \acp{pva} collected by the stacked vector $\overline{\RV{y}}^{(i)}_{m,n} \triangleq  [ \overline{\RV{y}}^{(1,i)\ist\T}_{m,n}  \ist\cdots\ist\ist\ist \overline{\RV{y}}^{(J,i)\ist\T}_{m,n} ]^{\T} $ are generated}. The unknown BS index $j$ itself is modeled as a discrete random variable $\overline{\rv{j}}^{(i)}_{m,n} \in \{1,\dots,  J\}$ with uniform prior \ac{pmf} $p(\overline{j}^{(i)}_{m,n}) \triangleq 1/J$. However, a new \ac{pva} can only correspond to one \ac{bs} resulting in the \ac{pdf}
\vspace*{-1mm}
\begin{align}
	&f\big( \overline{\V{p}}^{(j,i)}_{\text{va},m}|\V{x}_{n}^{(i)}, \overline{j}^{(i)}_{m,n} \big) =\rmv\begin{cases} 
		f_{\mathrm{n}}( \overline{\V{p}}^{(j,i)}_{\text{va},m}|\V{x}_{n}^{(i)}) , &\!\!\! j  \!=\rmv \overline{j}^{(i)}_{m,n} \\[0mm]
		0 , &\!\!\!  j  \!  \neq \rmv \overline{j}^{(i)}_{m,n} .
	\end{cases}\label{eq:newpot3}\\[-6mm]\nn
\end{align}
\Ac{fp} measurements generated by the \ac{ceda} are modeled as a Poisson process with mean $\mu_{\mathrm{fp}}$ and \ac{pdf} $f_{\mathrm{fp}}(\V{z}^{(i)}_{m,n}) = f_{\mathrm{fp}}({z_\mathrm{d}}^{(i)}_{m,n}) f_{\mathrm{fp}}({z_\mathrm{\varphi}}^{(i)}_{m,n}) f_{\mathrm{fp}}({z_\mathrm{\vartheta}}^{(i)}_{m,n})$, where the individual \ac{fp} \acp{lhf} are uniformly distributed over $[0,d_{\mathrm{max}}]$, $[-\pi,\pi)$, and $[-\pi,\pi)$, respectively.
We denote by $\overline{\RV{y}}^{(i)}_n \triangleq \big [ \ist \overline{\RV{y}}^{(i)\ist\T}_{1,n} \ist\cdots\ist \overline{\RV{y}}^{(i)\ist \T}_{ {\rv{M}_\text{bs}}^{(i)}_{n} \rmv\rmv\rmv,n} \ist \big]^{\T}\rmv\rmv\rmv$, the joint vector of all new \ac{pva} states.

\vspace{-2mm} 
\subsection{Legacy PVAs and Sequential Update} \label{sec:legacy_update3}

At time $n$, measurements are incorporated sequentially across \acp{mt} $i \rmv\in\rmv \{1,\dots,I\}$. Previously detected \acp{va}, i.e., \acp{va} that have been detected either at a previous time $n' \!<\rmv n$ or at the current time $n$ but at a previous \ac{mt} $i' \!<\rmv  i$, are therefore represented by legacy \ac{pva} states $\underline{\RV{y}}_{k,n}^{(i)}$, with $ \underline{\RV{y}}^{(1)}_{k,n} \triangleq \underline{\RV{y}}_{k,n}$. 
Additionally, each legacy \ac{pva} implicitly carries an index $j$ that identifies its corresponding \ac{bs}. It is given by the mapping
\begin{align} \label{eq:junderline}
	\underline{j}(k) \rmv\rmv\rmv=\rmv\rmv\rmv (k\rmv-\rmv1)\rmv\rmv\rmv\rmv\rmv\mod\rmv\rmv J+1 \, .
\end{align} 
New \acp{pva} become legacy \acp{pva} when the next measurements, either of the next \ac{mt} or at the next time instance, are taken into account. In particular, the \ac{va} represented by the new \ac{va} state $\overline{\RV{y}}^{(j,i')}_{m',n'}$ introduced due to measurement $m'$ of \ac{mt} $i'$ corresponding to \ac{bs} $j$ at $n' \rmv\rmv \leq \rmv\rmv n$ is represented by the legacy \ac{pva} state $\underline{\RV{y}}^{(i)}_{k,n}=[\underline{\RV{p}}^{(i)\ist\T}_{\text{va},m}  \iist \underline{\rv{r}}^{(i)}_{m}]^{\T}$ at time $n$, i.e., 
\vspace*{-3mm}
\begin{align} \label{eq:pva_index3}
	k = \rv{K}_{n'-1} + J\, \sum^{i' - 1}_{i'' = 1}  \rv{M}^{(i'')}_{n} + j\, m' \ist.\\[-7mm]\nn
\end{align}
The number of legacy \ac{pva} at time $n$, when the measurements of the next \ac{mt} $i$ are incorporated, is updated according to $\rv{K}^{(i)}_{n} = \rv{K}^{(i-1)}_{n-1} + J\, \rv{M}^{(i-1)}_{n}$, where $\rv{K}^{(1)}_{n} \rmv=\rmv \rv{K}_{n-1}$. Here, $\rv{K}^{(i)}_{n}$ is equal to the number of all measurements collected up to time $n$ and \ac{mt} $i \rmv-\rmv 1$. The vector of all legacy \ac{pva} states at time $n$ and up to \ac{mt} $i$ can now be written as $\underline{\RV{y}}^{(i)}_{n} = \big[\underline{\RV{y}}^{(i-1) \T}_{n} \ist\ist\ist \overline{\RV{y}}^{(i-1) \T}_{n} \big]^{\T}\rmv\rmv$ and the current vector containing all states as ${\RV{y}}^{(i-1)}_{n} = \big[\underline{\RV{y}}^{(i-1) \T}_{n} \ist\ist\ist \overline{\RV{y}}^{(i-1) \T}_{n} \big]^{\T}\rmv\rmv$. 

Let us denote by $\underline{\RV{y}}^{(1)}_n \rmv\triangleq\rmv \big[ \underline{\RV{y}}^{T}_{1,n} \rmv\cdots\ist \underline{\RV{y}}^{\T}_{\rv{K}_{n\rmv-\rmv1},n} \big]^{\T}\rmv\rmv$, the vector of all legacy \ac{pva} states before any measurements at time $n$ have been incorporated. After the measurements of all \acp{mt} $i \in \{1,\dots,I\}$ have been incorporated at time $n$, the total number of \ac{pva} states is 
\vspace*{-2mm}
\begin{align} 
	\rv{K}_n = \rv{K}_{n-1} + J \, \sum_{i=1}^{I}  \rv{M}^{(i)}_{n} = \rv{K}^{(I)}_{n} + \rv{M}^{(I)}_{n} \label{eq:nrpvatotal}\\[-8mm]\nn
\end{align}
and the vector of all \ac{pva} states at time $n$ is given by $\RV{y}_{n} \!=\rmv \big[\underline{\RV{y}}^{(I)\ist \T}_{n} \ist\ist \overline{\RV{y}}^{(I) \T}_{n} \big]^\T$.

\vspace{-2mm}
\subsection{State Evolution} 
\vspace{-1mm}
\label{sec:state_evolution} 
Legacy \acp{pva} states $\underline{\RV{y}}_{k,n}$ and the \ac{mt} states $\RV{x}^{(i)}_{n}$ are assumed to evolve independently across time according to state-transition \acp{pdf} $f(\V{x}^{(i)}_{n}|\V{x}^{(i)}_{n-1})$, $f(b_{\text{bs},n}^{(j)}| b_{\text{bs},n-1}^{(j)})$, and $f\big(\underline{\V{y}}_{k,n} \big| \V{y}_{k, n-1}\big)$, respectively. In particular, if \ac{pva} $k$ exists at time $n \rmv-\! 1$, i.e., $r_{k,n-1} \!=\! 1$, it either disappears, i.e., $\overline{r}_{k,n} \!=\rmv 0$, or survives, i.e., $\overline{r}_{k,n} \!=\! 1$; in the latter case, it becomes a legacy \ac{pva} at time $n$. The probability of survival is denoted by $p_\mathrm{s}$. In case of survival, its position remains unchanged, i.e., the state-transition \ac{pdf} of the \ac{va} positions $\underline{\RV{p}}_{\text{va},k}$ is given by $f\big(\underline{\V{p}}_{\text{va},k}  \ist \big| \ist \V{p}_{\text{va},k} \big) = \delta \big(\underline{\V{p}}_{\text{va},k} \rmv - \ist \V{p}_{\text{va},k} \big)$. 
Therefore, $f\big(\underline{\V{p}}_{\text{va},k}\rmv,\underline{r}_{k,n} \ist \big| \ist \V{p}_{\text{va},k} , r_{k,n-1} \big) $ for $r_{k,n-1} \rmv=\rmv 1$ is obtained as  
\vspace*{-2mm}
\begin{align}
	f\big(\underline{\V{p}}_{\text{va},k},\underline{r}_{k,n} \ist \big| \ist \V{p}_{\text{va},k} , r_{k,n-1} = 1\big) & \nn \\
	& \hspace{-30mm} =\rmv\begin{cases} 
		(1 \!-\rmv p_\mathrm{s}) \ist f_\text{d}\big(\underline{\V{p}}_{\text{va},k} \big) , &\!\!\! \underline{r}_{k,n} \!=\rmv 0 \\[0mm]
		p_\mathrm{s} \ist\ist   \delta \big(\underline{\V{p}}_{\text{va},k} \rmv - \ist \V{p}_{\text{va},k} \big) , &\!\!\! \underline{r}_{k,n} \!=\! 1.
	\end{cases}\label{eq:stmpvarone}\\[-7mm]\nn
\end{align}
If \ac{va} $k$ does not exist at time $n \rmv-\! 1$, i.e., $r_{k,n-1} \!=\! 0$, it cannot exist as a legacy \ac{pva} at time $n$ either, thus we get
\vspace*{-1mm}
\begin{align}
	&f\big(\underline{\V{p}}_{\text{va},k},\underline{r}_{k,n} \big| \V{p}_{\text{va},k} , 0\big) =\rmv\begin{cases} 
		f_\text{d}\big( \underline{\V{p}}_{\text{va},k} \big) , &\!\!\! \underline{r}_{k,n} \!=\rmv 0 \\[0mm]
		0 , &\!\!\! \underline{r}_{k,n} \!=\! 1.
	\end{cases}\label{eq:stmpvarzero}\\[-7mm]\nn
\end{align}
To account for the sequential update, we define $f^{(i)}\big(\underline{\V{y}}^{(i)}_{k,n} \big| {\V{y}}^{(i-1)}_{k, n}\big)$ for $i \geq 2$ and $\underline{r}^{(i)}_{k-1,n} = 1$ as
\vspace*{-1mm}
\begin{align}
	\hspace*{-4mm}f^{(i)}\big(\underline{\V{p}}^{(i)}_{\text{va},k},\underline{r}^{(i)}_{k,n} \ist \big| {\ist {\V{p}}^{(i-1)}_{\text{va},k} , 1}\big) 
	\rmv=\rmv\rmv\begin{cases} 
		f_\text{d}\big(\underline{\V{p}}^{(i)}_{\text{va},k} \big) , &\!\!\! \underline{r}^{(i)}_{k,n} \!=\rmv 0 \\[0mm]
		\delta \big(\underline{\V{p}}^{(i)}_{\text{va},k} \rmv - \ist {\V{p}}^{(i-1)}_{\text{va},k} \big) , &\!\!\! \underline{r}^{(i)}_{k,n} \!=\! 1
	\end{cases}\label{eq:stmpvaonepas}\\[-7mm]\nn
\end{align}
and for $\underline{r}^{(i)}_{k-1,n} = 0$
\vspace*{-2mm}
\begin{align}
	f^{(i)}\big(\underline{\V{p}}^{(i)}_{\text{va},k},\underline{r}^{(i)}_{k,n} \big| {{\V{p}}^{(i-1)}_{\text{va},k} , 0}\big) 
	=\rmv\begin{cases} 
		f_\text{d}\big( \underline{\V{p}}^{(i)}_{\text{va},k} \big) , &\!\!\! \underline{r}^{(i)}_{k,n} \!=\! 0 \\[0mm]
		0 , &\!\!\! \underline{r}^{(i)}_{k,n} \!=\! 1.
	\end{cases}\label{eq:stmpvazeropas}\\[-7mm]\nn
\end{align}
The initial prior \ac{pdf} at $n \rmv=\rmv 0$ are given as $f\big(\V{y}^{(i)}_{k,0} \big)$, $k = \big\{1,\dots,K^{(i)}_0\big\}$ and  $f(\V{x}^{(i)}_{0})$, where $\RV{y}^{(i)}_{k,0}$ typically is an empty vector. All (legacy and new) \ac{pva} states and all \ac{mt} states up to time $n$ are denoted as $\RV{y}_{n} \!=\rmv \big[\RV{y}^{(1)\ist \T}_{n} \cdots\ist \RV{y}^{(i) \T}_{n} \big]^\T$ and $\RV{y}_{0:n} \triangleq \big[\RV{y}^{\T}_{0} \cdots\ist \RV{y}^{\T}_{n} \big]^{\T}\!$ and $\RV{x}_{n} = [\RV{x}_{1,n}^\T \ist\cdots \RV{x}_{I}^\T]$ and $\RV{x}_{0:n} = [\RV{x}_{0}^\T \ist\cdots \RV{x}_{n}^\T]$, respectively.
To model the sequential update of the \ac{bs} biases, we define the factors $f^{(i)}(b_{\text{bs},n}^{(j,i)} | b_{\text{bs},n}^{(j,i-1)})$ for $i \geq 2$ as $f^{(i)}(b_{\text{bs},n}^{(j,i)} | b_{\text{bs},n}^{(j,i-1)})=\delta(b_{\text{bs},n}^{(j,i)}-b_{\text{bs},n}^{(j,i-1)})$ with $b_{\text{bs},n}^{(j)} \triangleq b_{\text{bs},n}^{(j,I)}$. For $i=1$, we use $f^{(1)}(b_{\text{bs},n}^{(j,1)} | b_{\text{bs},n}^{(j,0)})$ with $b_{\text{bs},n}^{(j,0)} \triangleq b_{\text{bs},n-1}^{(j)}$.

\vspace{-2mm} 
\subsection{Data Association}

Data association between measurements and \acp{pva} is described in a redundant formulation by the \ac{pva}/\ac{pa}-oriented association vector $ \underline{\RV{a}}_{n}^{(i)} \triangleq [\underline{\rv{a}}_{1,n}^{(i)} \ist \cdots \ist  \underline{\rv{a}}_{J,n}^{(i)} \iist \underline{\rv{a}}_{J+1,n}^{(i)} \ist \cdots \ist \underline{\rv{a}}_{J+\rv{K}^{(i)}_{n},n}^{(i)}]^{\mathrm{T}} $ and by the measurement-oriented association vector $ \overline{\RV{a}}_{n}^{(i)} \triangleq [\overline{\rv{a}}_{1,n}^{(i)} \ist \cdots \ist \overline{\rv{a}}_{{\rv{M}}_n^{(i)},n}^{(i)}]^{\mathrm{T}} $. See \cite{WilLau:J14,MeyKroWilLauHlaBraWin:J18,LeiMeyHlaWitTufWin:J19,LeiWieVenWit:Asilomar2024} for details. 

\begin{figure*}[!]
	\vspace*{-5mm}
	\scalebox{0.8}{
		\parbox{1.2\linewidth}{
			\begin{align}
				&f\big( \V{y}_{0:n}, \V{x}_{0:n}, \underline{\V{a}}_{1:n},\overline{\V{a}}_{1:n},\overline{\V{j}}_{1:n} \ist \big| \ist \V{z}_{1:n}\big)\nn\\[-1mm]
				&\hspace{0mm} \propto\underbrace{\bigg( \prod^{N_{\text{MT}}}_{i = 1}\ist f(\V{x}^{(i)}_{0}) \ist \prod^{K_0}_{k = 1} \ist f(\V{y}_{k,0})  \bigg)}_{\footnotesize \text{MT and VA states initial prior PDFs}}\prod^{n}_{n' = 1}  \underbrace{\bigg( \prod^{N_{\text{MT}}}_{i = 1}\ist f\big(\V{x}^{(i)}_{n'} | \V{x}^{(i)}_{n'-1} \big)  }_{\footnotesize \text{MT state prediction}} \underbrace{\bigg(\prod^{J}_{j = 1} \ist f\big(b_{\text{bs},n}^{(j,i)}| b_{\text{bs},n}^{(j,i-1)}\big) \underline{q}_\text{BS}\big(\V{x}^{(i)}_{n}, {b}^{(j,i)}_{\text{bs},n}, \underline{a}^{(i)}_{j,n}; {\V{z}_\text{bs}}^{(i)}_{n} \big) \prod^{M^{(j)}_{n'}}_{m' = 1} \rmv\rmv \Psi\big(\underline{a}^{(i)}_{j,n'},\overline{a}^{(i)}_{m',n'} \big) \bigg)}_{\footnotesize \text{Factors and state prediction related to \acp{bs}}} 		\underbrace{\bigg( \prod^{K_{n'\rmv-\rmv1}}_{k' = 1} f \big(\underline{\V{y}}_{k'\rmv\rmv,n'} | \V{y}_{k'\rmv\rmv,n'-1}\big) \rmv \bigg)}_{\footnotesize \text{Legacy PVA states prediction}}
				\nn\\[-1mm]	
				& \hspace{0mm} \times\ 		
				\underbrace{ \bigg(\prod^{N_{\text{MT}}}_{i = 2}\ist \prod^{K^{(i)}_{n'}}_{k = 1}\ist f^{(i)}\big(\underline{\V{y}}^{(i)}_{k,n'} \big| \underline{\V{y}}^{(i-1)}_{k, n'}\big) \bigg)}_{\footnotesize  \text{Legacy PVA states transition factors}} 
				\underbrace{\prod^{N_{\text{MT}}}_{i = 1} \prod^{K^{(i)}_{n'}}_{k = 1} \bigg( \ist \underline{q}\big( \underline{\V{y}}^{(i)}_{k,n'}\rmv,  \V{x}^{(i)}_{n'}, {b}^{(\underline{j}(k),i)}_{\text{bs},n}, \underline{a}^{(i)}_{k+J,n'} ; {\V{z}_\text{bs}}^{(i)}_{n'} \big) \prod^{{M_\text{bs}}^{(i)}_{n'}}_{m = 1} \rmv\rmv \Psi\big(\underline{a}^{(i)}_{k,n'},\overline{a}^{(i)}_{m,n'} \big)\bigg)}_{\footnotesize  \text{Legacy PVA states related factors}} \underbrace{ \prod^{{M_\text{bs}}^{(i)}_{n'}}_{m = 1} \overline{q}\big( \overline{\V{y}}^{(i)}_{m,n'}, \V{x}^{(i)}_{n'}, \bm{b}^{(i)}_{\text{bs},n}, \overline{j}^{(i)}_{m,n}, \overline{a}^{(i)}_{m,n'} ; {\V{z}_\text{bs}}^{(i)}_{n'} \big) \rmv }_{\footnotesize  \text{New PVA states prior PDF and related factors}}\label{eq:jointpostpdf}
			\end{align}
		}
	}
	\vspace*{-7mm}
\end{figure*}

\vspace{-2mm} 
\section{Factor Graph and Sum-Product Algorithm} \label{sec:factor_graph}
\subsubsection{Joint Posterior PDF and Factor Graph}

Using Bayes' rule and independence assumptions related to the state-transition \acp{pdf}, the prior \acp{pdf}, and the likelihood model (for details please see \cite{MeyKroWilLauHlaBraWin:J18,LeiMeyHlaWitTufWin:J19,LeiVenTeaMey:TSP2023}), and for fixed (observed) measurements $\V{z}_{1:n}$ (the numbers of measurements ${M}^{(i)}_{n}$ are fixed and not random anymore) the joint posterior \ac{pdf} of $\RV{y}_{0:n}$, $\RV{x}_{0:n}$, $\underline{\RV{a}}_{1:n}$, and $\overline{\RV{a}}_{1:n}$, conditioned on $\V{z}_{1:n}$ is given by \eqref{eq:jointpostpdf}, where the individual \ac{lhf}-related factors are respectively $\underline{q}_\text{BS}\big( \V{x}_{n}^{(i)}, {b}^{(j,i)}_{\text{bs},n},\underline{a}^{(i)}_{j,n}; {\V{z}_\text{bs}}^{(i)}_{n} \big)$, $\underline{q}\big( \underline{\V{y}}^{(i)}_{k,n}, \V{x}_{n}^{(i)}, {b}^{(\underline{j}(k),i)}_{\text{bs},n}\rmv\rmv\rmv,\underline{a}^{(i)}_{k+J,n}; {\V{z}_\text{bs}}^{(i)}_{n} \big)$ 
and $\overline{q}\big( \overline{\V{y}}^{(i)}_{m,n}, \V{x}^{(i)}_{n},\bm{b}^{(i)}_{\text{bs},n},  $ $  \overline{j}^{(i)}_{m,n}, \overline{a}^{(i)}_{m,n}; {\V{z}_\text{bs}}^{(i)}_{n} \big)$ that will be discussed next. 

The \textit{pseudo \acp{lhf}} related to \ac{bs} $j$ and \ac{mt} $i$ are given by
\begin{align}
	&\underline{q}_\text{BS}\big(\V{x}_{n}^{(i)}, {b}^{(j,i)}_{\text{bs},n}, \underline{a}^{(i)}_{j,n}; {\V{z}_\text{bs}}^{(i)}_{n} \big)\nn\\ &
	\hspace*{10mm}\triangleq \begin{cases}
		\displaystyle \ist \frac{ 	p_{\mathrm{d}}  f\big( {\V{z}_\text{bs}}^{(i)}_{m,n} \big|\ist \V{x}^{(i)}_{n}, {b}^{(j,i)}_{\text{bs},n} \big)}{ \mu_{\mathrm{fp}} \ist f_{\mathrm{fp}}\big( {\V{z}_\text{bs}}^{(i)}_{m,n} \big)} \ist, 
		& \!\!\rmv \underline{a}^{(i)}_{j,n} \!=\rmv m \\[3.5mm]
		1 \!-\rmv 	p_{\mathrm{d}}\ist, & \!\!\rmv \underline{a}^{(i)}_{j,n} \!=\rmv 0   
	\end{cases} \label{eq:bslhf}
\end{align}

The \textit{pseudo \ac{lhf}} for legacy \ac{pva} $k$ and \ac{mt} $i$ is given by
\begin{align}
	&\underline{q}\big(\underline{\V{p}}^{(i)}_{k,\text{va}},\underline{r}^{(i)}_{k,n}=1, \V{x}_{n}^{(i)}, {b}^{(\underline{j}(k),i)}_{\text{bs},n},\underline{a}^{(i)}_{k+J,n}; {\V{z}_\text{bs}}^{(i)}_{n} \big)\nn\\ &\triangleq \begin{cases}
		\displaystyle \ist \frac{ 	p_{\mathrm{d}}  f\big( {\V{z}_\text{bs}}^{(i)}_{m,n} \big|\ist \V{x}^{(i)}_{n}, \underline{\V{p}}^{(i)}_{k,\text{va}}\rmv, {b}^{(\underline{j}(k),i)}_{\text{bs},n} \big)}{ \mu_{\mathrm{fp}} \ist f_{\mathrm{fp}}\big( {\V{z}_\text{bs}}^{(i)}_{m,n} \big)} \ist, 
		& \!\!\rmv \underline{a}^{(i)}_{k+J,n} \!=\rmv m \\[3.5mm]
		1 \!-\rmv 	p_{\mathrm{d}}\ist, & \!\!\rmv \underline{a}^{(i)}_{k+J,n} \!=\rmv 0   
	\end{cases} \label{eq:lvalhf}
\end{align}
and $\underline{q}\big( \underline{\V{x}}^{(i)}_{k,n},\underline{r}^{(i)}_{k,n}=0, \V{x}_{n}^{(i)}, {b}^{(\underline{j}(k),i)}_{\text{bs},n},\underline{a}^{(i)}_{k+J,n}; {\V{z}_\text{bs}}^{(i)}_{n} \big)   \rmv\triangleq\rmv {\delta_{\underline{a}^{(i)}_{k,n}}}$. 

\begin{figure}[t]
	\centering
	\scalebox{0.60}{\includegraphics{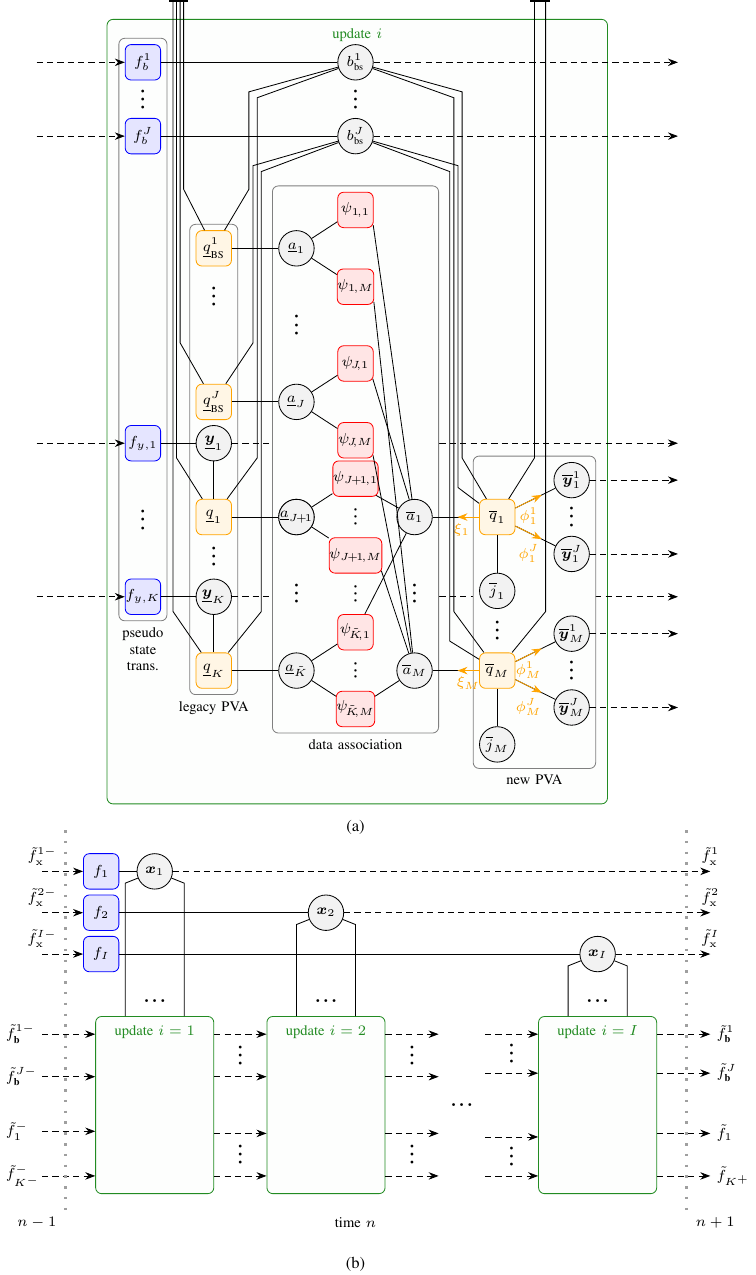}}
	\vspace*{-3mm}
	\caption{Factor graph corresponding to the factorization shown in \eqref{eq:jointpostpdf}. 
		Dashed arrows represent messages that are only passed in one direction. The detailed graph in (a) represents the green subgraphs of the overall factor graph in (b). The following short notation is used. 
		Variables and variable nodes: 
		$\V{x}_i \triangleq \V{x}^{(i)}_{n}$,
		$\underline{\V{y}}^{i}_k \triangleq \underline{\V{y}}^{(i)}_{k,n}$,
		$\overline{\V{y}}^{j}_m \triangleq \overline{\V{y}}^{(j,i)}_{m,n}$,
		$\underline{a}_k \triangleq \underline{a}^{(i)}_k$,
		$\overline{a}_m \triangleq \overline{a}^{(i)}_m$,	
		$b_{\text{bs}}^{j} \triangleq b_{\text{bs},n}^{(j)} $,	
		$ \overline{j}_m \triangleq \overline{j}^{(i)}_{m,n}$,
		$K^- \triangleq K_{n-1}$, 
		$K \triangleq K^{(i)}_{n}$
		$\tilde{K} = J+ K^{(i)}_{n}$, and
		$K^+ \triangleq K_{n}$.
		Factor node: 
		$f_{i} \triangleq f(\V{x}^{(i)}_{n}|\V{x}^{(i)}_{n-1})$,
		$f_{y,k} \triangleq f(\underline{\V{y}}_{k,n} | \V{y}_{k,n-1})$ for $i = 1$,
		$f_{y,k} \triangleq f(\underline{\V{y}}^{(i)}_{k,n} | \V{y}^{(i-1)}_{k,n-1})$ for $i > 1$,
		$f^{j}_{\text{b}} \triangleq f^{(i)}(b_{\text{bs},n}^{(j,i)}| b_{\text{bs},n}^{(j,i-1)})$, and
		$\psi_{ k,m} \triangleq \Psi(\underline{a}^{(i)}_{k,n},\overline{a}^{(i)}_{m,n}) $.
		Messages: 
		$\tilde{f}_\text{x}^{i} \triangleq f(\V{x}^{(i)}_{n}|\V{z}_{1:n})$, 
		$\tilde{f}_\text{b}^{j} \triangleq f(b_{\text{bs},n}^{(j)}|\V{z}_{1:n})$, and
		$\tilde{f}_k \triangleq f(\V{y}_{k}|\V{z}_{1:n})$, where $\V{y}_{k}$ represents an entry of $\RV{y}_{n} \!=\rmv \big[\underline{\RV{y}}^{(I)\ist \T}_{n} \ist\ist \overline{\RV{y}}^{(I) \T}_{n} \big]^\T$.
		The dashed arrows indicate messages representing MT and PVA beliefs of time
		$n-1$ and $n+1$, which are only propagated forward in time and the minus sign ``$-$'' indicates beliefs from the previous time $n-1$.
	}	 
	\label{fig:factorGraph}
	\vspace*{-5mm}
\end{figure}

The \textit{pseudo \ac{lhf}} for new \ac{pva} $m$ and \ac{mt} $i$ is given by
\vspace*{-1mm}
\begin{align}
	& \overline{q}\big(\overline{\V{y}}^{(i)}_{m,n}, \V{x}_{n}^{(i)}, \bm{b}^{i}_{\text{bs},n},  \overline{j}^{(i)}_{m,n}, \overline{a}^{(i)}_{m,n}; {\V{z}_\text{bs}}^{(i)}_{n} \big) \nn\\
	&\hspace*{5mm}\triangleq \begin{cases}
		0    \ist, 
		& \hspace{-1mm} j \!\neq\rmv \overline{j}^{(i)}_{m,n}   \\[1mm] %
		\overline{q}\big( \overline{\V{y}}^{(i)}_{m,n}, \V{x}_{n}^{(i)}, {b}^{(j,i)}_{\text{bs},n}, \overline{a}^{(i)}_{m,n}; {\V{z}_\text{bs}}^{(i)}_{n} \big) \ist,  & \hspace{-1mm} j  \!=\rmv \overline{j}^{(i)}_{m,n}   
	\end{cases}  \label{eq:nvalhf}\\[-6mm]\nn
\end{align}
with
\vspace*{-1mm}
\begin{align}
	&\hspace{-1mm}\overline{q}\big( \overline{\V{p}}^{(j,i)}_{\text{va},m}, \overline{r}^{(j,i)}_{m,n}=1, \V{x}_{n}^{(i)}, {b}^{(j,i)}_{\text{bs},n}, \overline{a}^{(i)}_{m,n}; {\V{z}_\text{bs}}^{(i)}_{n} \big)\nn\\ 
	&\hspace{-1mm}\triangleq \begin{cases}
		0    \ist, 
		& \hspace{-1mm} \overline{a}^{(i)}_{m,n}  \rmv\rmv=\rmv k   \\[1mm] %
		\frac{ \mu_{\mathrm{n}}f_{\mathrm{n}}( \overline{\V{p}}^{(j,i)}_{\text{va},m}\ist | \ist\V{x}_{n}^{(i)}) f({\V{z}_\text{bs}}^{(i)}_{m,n} \ist | \V{x}_{n}^{(i)}, \overline{\V{p}}^{(j,i)}_{\text{va},m}, {b}^{(j,i)}_{\text{bs},n})}{J\, \mu_{\mathrm{fp}}  f_{\mathrm{fp}}({\V{z}_\text{bs}}^{(i)}_{m,n} )} \ist,  & \hspace{-1mm} \overline{a}^{(i)}_{m,n} \rmv=\rmv 0 
	\end{cases}  \label{eq:factorvNewPVAs}\\[-6mm]\nn
\end{align}
and $\overline{q}\big( \overline{\V{p}}^{(j,i)}_{\text{va},m},\overline{r}^{(j,i)}_{m,n}=0, \V{x}_{n}^{(i)}, {b}^{(j,i)}_{\text{bs},n}, \overline{a}^{(i)}_{m,n}; {\V{z}_\text{bs}}^{(i)}_{n} \big) \rmv\triangleq\rmv 	f_\text{d}\big( \overline{\V{p}}^{(j,i)}_{\text{va},m}\big)$, respectively. Finally, the binary \textit{indicator functions} that check consistency for any pair $(\underline{a}^{(i)}_{k,n},\overline{a}^{(i)}_{m,n})$ of \ac{pva}/\ac{pa}-oriented and measurement-oriented association variable at time $n$, \vspace{-.4mm} read
\begin{align}
	&\hspace{-2mm}\Psi(\underline{a}^{(i)}_{k,n},\overline{a}^{(i)}_{m,n}) \nn\\
	&\hspace{-2mm}\triangleq \begin{cases} 
		0, & \underline{a}^{(i)}_{k,n} =  m,\overline{a}^{(i)}_{m,n} \neq k \text{ or } \underline{a}^{(i)}_{k,n} \neq  m, \overline{a}^{(i)}_{m,n} = k  \\[.1mm]
		1 \ist, & \text{otherwise}. 
	\end{cases} \\[-6mm]
	\nn
\end{align}
The factor graph equivalently representing factorization \eqref{eq:jointpostpdf} is shown in Fig.~\ref{fig:factorGraph}.

\subsubsection{Confirmation of PVAs and State Estimation}\label{sec:probFormulation3}
We estimate the \ac{mt} states $\RV{x}^{(i)}_{n}$ (including $\rv{b}_{\text{mt}\, n}^{(i)}$), the \ac{bs} biases $\rv{b}_{\text{bs}\, n}^{(j)}$, and the \ac{pva} states $\RV{p}_{\text{va},k}$ from all measurements $\V{z}_{1:n}$ up to time~$n$ using the respective \ac{mmse} estimates \cite[Ch.~4]{Kay1993} $\hat{{\V{x}}}^{(i)}_{n} = [\hat{\tilde{\V{x}}}^{(i)}_{n}\iist\hat{\V{v}}^{(i)}_{n}\iist\hat{b}_{\text{mt}, n}^{(i)}]$ and $	\hat{b}_{\text{bs}\, n}^{(j)}$, which are calculated by the expected values of the respective marginal posterior distributions $f( \V{x}^{(i)}_{n}|\V{z}_{1:n}) $, $f({b}_{\text{bs}, n}^{(j)}|\V{z}_{1:n})$. 
Detection of the \acp{pva} $k\!\in\!\{1,\dots,K_n\}$ relies on the posterior existence probabilities $p(r_{k,n}\!=\!1|\V{z}_{1:n}) = \int f(\V{p}_{\text{va},k}, r_{k,n}\!=\!1|\V{z}_{1:n})\, \mathrm{d}\V{p}_{\text{va},k}$ and the conditional posteriors $f(\V{p}_{\text{va},k}|r_{k,n}\!=\!1,\V{z}_{1:n}) = f(\V{p}_{\text{va},k}, r_{k,n}\!=\!1|\V{z}_{1:n})/p(r_{k,n}\!=\!1|\V{z}_{1:n})$. A \ac{pva} is confirmed if $p(r_{k,n}\!=\!1|\V{z}_{1:n})>p_{\mathrm{cf}}$ \cite{Kay1998}.

Since introducing new \acp{pva} causes $K_n$ to grow indefinitely, a suboptimal pruning step is applied. A \acp{pva} is pruned if $p(r_{k,n}\!=\!1|\V{z}_{1:n})<p_{\text{pr}}$, except for $k\!=\!1$ of each \ac{bs}, which is always retained. Note that before pruining $\underline{j}(k)$ needs to be saved, since \eqref{eq:junderline} becomes invalid. For existing \acp{pva}, we determine the MMSE estimate $\hat{\V{p}}_{\text{va},k}$ by calculating the expected value of $f(\V{p}_{\text{va},k}|r_{k,n}\!=\!1,\V{z}_{1:n})$. 
Since direct marginalization of \eqref{eq:jointpostpdf} is intractable, sequential message passing via the SPA on a factor graph \cite{KscFreLoe:TIT2001} provides efficient approximations (``beliefs'') of all marginal posteriors  $f(\V{x}^{(i)}_{n}|\V{z}_{1:n})$, $f(b_{\text{bs},n}^{(j)}|\V{z}_{1:n})$, $f(\V{p}_{\text{va},k}|r_{k,n}\!=\!1,\V{z}_{1:n})$, and $p(r_{k,n}|\V{z}_{1:n})$, for all \ac{mt} states $\V{x}^{(i)}_{n}$, \ac{bs} biases $b_{\text{bs},n}^{(j)}$, and \acp{pva} $k$.
The runtime of the proposed algorithm scales linearly in the number of \acp{pva}, measurements, \acp{mt} and particles, i.e, $\mathcal{O}(J K_{n-1}^{(i)} M_n^{(i)} P)$. The number of \acp{pva}, which increases over time $n$ and MT index $i$ according to \eqref{eq:nrpvatotal}, can become large. Compared to state-of-the-art cooperative \ac{mpslam} \cite{LeiWieVenWit:Asilomar2024} the proposed algorithm involves an additional factor $J$, since a new \ac{pva} is generated for each anchor. This is counteracted by pruning.

\subsubsection{Selected SPA Messages}

A complete derivation of all \ac{spa} messages is omitted due to page limitation; only the key messages relevant to new \acp{pva} are presented (see \cite{LeiMeyHlaWitTufWin:J19,LeiVenTeaMey:TSP2023}). 

\paragraph{Measurement Evaluation for New PVAs}\label{sec:xiMess3} The messages $\xi\big(\overline{a}^{(i)}_{m,n}\big)$ sent from the factor node $\overline{q}\big(\overline{\V{y}}^{(i)}_{m,n}, \V{x}_{n}^{(i)}, \bm{b}^{i}_{\text{bs},n}, $ $ \overline{j}^{(i)}_{m,n}, \overline{a}^{(i)}_{m,n}; {\V{z}_\text{bs}}^{(i)}_{n} \big)$ to the variable nodes corresponding to the measurement-oriented association variables $\overline{a}^{(i)}_{m,n}$ are given by
\begin{align}
	&\xi\big(\overline{a}^{(i)}_{m,n}\big) =\rmv\rmv \sum_{\overline{j}^{(i)}_{m,n}=1}^{J} \multisum_{\overline{r}^{(j,i)}_{m,n} \ist\ist\in\ist\ist \overline{\V{r}}^{(i)}_{m,n}} \int\!\!\!\int \rmv\tilde{f}^{(i)}(\V{x}_{n}^{(i)}) \ist \overline{q}\big(\overline{\V{p}}^{(i)}_{\text{va},m}, \overline{\V{r}}^{(i)}_{m,n}
	, \nn\\
	&\hspace*{15mm}\V{x}_{n}^{(i)}, \bm{b}^{i}_{\text{bs},n},  \overline{j}^{(i)}_{m,n}, \overline{a}^{(i)}_{m,n}; {\V{z}_\text{bs}}^{(i)}_{n} \big) \, \mathrm{d}\V{x}_{n}^{(i)} \ist \mathrm{d}\overline{\V{p}}^{(i)}_{\text{va},m} 
	\label{eq:bp_measevalutionNF1_phase3}\\[-6mm]
	\nn
\end{align}
where the new \ac{pva} state $\overline{\RV{y}}^{(i)}_{m,n} = [\overline{\V{p}}^{(i)}_{\text{va},m}\iist\overline{\V{r}}^{(i)}_{m,n}]^{\T}$ is replaced by its entries. Using the expression of $\overline{q}\big(\overline{\V{y}}^{(i)}_{m,n}, \V{x}_{n}^{(i)}, \bm{b}^{i}_{\text{bs},n}, $ $ \overline{j}^{(i)}_{m,n}, \overline{a}^{(i)}_{m,n}; {\V{z}_\text{bs}}^{(i)}_{n} \big)$ in \eqref{eq:factorvNewPVAs}, Eq.\ \eqref{eq:bp_measevalutionNF1_phase3} simplifies to $\xi\big(\overline{a}^{(i)}_{m,n}\big) \!=\! 1$
for $\overline{a}^{(i)}_{m,n} \!\in\! \Set{K}_{n}$, and for $\overline{a}^{(i)}_{m,n} \!=\rmv 0$ it becomes
\vspace*{-3mm}
\begin{align}
	&\hspace*{-3mm}\xi\big(\overline{a}^{(i)}_{m,n}\big) = 1 + \frac{ \mu_{\text{n}} }{J \, \mu_\mathrm{fp} \ist f_{\mathrm{fp}}( {\V{z}_\text{bs}}^{(i)}_{n} )} \sum_{j=1}^{J}
	\int\!\!\!\int\rmv f_{\text{n}}(\overline{\V{p}}^{(j,i)}_{\text{va},m}| \V{x}_{n}^{(i)} ) \nn\\
	&\hspace*{0mm} \times  f({\V{z}_\text{bs}}^{(i)}_{m,n} \ist | \V{x}_{n}^{(i)}, \overline{\V{p}}^{(j,i)}_{\text{va},m}, {b}^{(j,i)}_{\text{bs},n}) \, \tilde{f}^{(i)}(\V{x}_{n}^{(i)}) \ist \mathrm{d}\V{x}_{n}^{(i)} \ist \mathrm{d}\overline{\V{p}}^{(j,i)}_{\text{va},m}\ist.
	\label{eq:bp_measevalutionNF_phase3}\\[-6mm]\nn
\end{align}

\paragraph{Measurement Update for New PVAs} 

Finally, the messages $\phi\big(\overline{\V{y}}^{(j,i)}_{m,n}\big) \triangleq \phi\big(\overline{\V{x}}^{(j,i)}_{m,n},\overline{{r}}_{m,n}^{(j,i)}\big)$ sent to each new \ac{pva} variable node are obtained as
\vspace*{-1mm}
\begin{align}
	\label{eq:measurementUpdateNew0_phase3}
	&\phi\big(\overline{\V{x}}^{(j,i)}_{m,n}, \overline{{r}}^{(j,i)}_{m,n}\big) \nn\\
	&\hspace*{2mm}= \ist\int\!\overline{q}\big(\overline{\V{p}}^{(i)}_{\text{va},m},  \overline{\V{r}}^{(i)}_{m,n}, \overline{a}^{(i)}_{m,n}=0, \V{x}_{n}^{(i)} , j^{(i)}_{m,n} = j ; {\V{z}_\text{bs}}^{(i)}_{n} \big) \nn\\ 
	&\hspace*{10mm}\times \tilde{f}^{(i)}(\V{x}_{n}^{(i)})  \, \mathrm{d}\V{x}_{n}^{(i)} \,\varsigma\big( \overline{a}^{(i)}_{m,n}=0\big)\\[-7mm]\nn 
\end{align}
resulting in 
\vspace*{-1mm}
\begin{align}
	\label{eq:measurementUpdateNew1_phase3}
	\phi\big(\overline{\V{x}}^{(j,i)}_{m,n}, 1\big) \big) &= \int\! \overline{q}_j\big(\overline{\V{p}}^{(j,i)}_{\text{va},m}, 1, 0, \V{x}_{n}^{(i)} ; {\V{z}_\text{bs}}^{(i)}_{n} \big) \nn\\
	&\hspace*{5mm} \times\tilde{f}^{(i)}(\V{x}_{n}^{(i)})\, \mathrm{d}\V{x}_{n}^{(i)} \,\varsigma\big( \overline{a}^{(j,i)}_{m,n}=0\big) \\[2mm] 
	\phi\big(\overline{\V{x}}^{(j,i)}_{m,n},0\big) &\triangleq\ist \phi_{m,n}^{(j,i)}  =\rmv \sum_{\overline{a}_{m,n}^{(i)} \in  \Set{K}^{(i)}_{n}} \!\! \varsigma\big( \overline{a}_{m,n}^{(i)}\big) \ist.
	\label{eq:measurementUpdateNew2_phase3}\\[-6mm]\nn
\end{align}

\vspace{-1mm}
\section{Results}\label{sec:results}

We consider a simple indoor scenario as shown in Fig.~\ref{fig:converging_pvas}. 
The scenario comprises four \acp{bs} and 4 reflective surfaces, i.e., $4$ \acp{va} per BS. Three \acp{mt} move along tracks which are observed for $400$ time instances $n$ with observation period $\Delta T = 1\,$s. 
Measurements are generated according to the system model in Section~\ref{sec:system_model}. The \ac{snr} is set to {$40\,\mathrm{dB}$} at a \ac{los} distance of $1\,$m. The amplitudes of the \acp{mpc} (including the \ac{los} component) are calculated using free-space path loss and an additional attenuation of {$3\,\mathrm{dB}$} for each reflection at a surface. The detection threshold is $\gamma = 8\,$dB and $p_\text{d} = 0.8$. False positive measurements are generated according to the model in Section~\ref{sec:signal_model} with a mean number of \acp{fp} $\mu_\text{fp} = 5$. For the calculation of the measurement variances, we assume a system bandwidth of $B=100\,\mathrm{MHz}$ at a center frequency of $f_\mathrm{c} = 6\,\mathrm{GHz}$. The arrays employed at \acp{mt} and \acp{pa} are of identical geometry with $H = H^{(j)}=4$ antenna elements forming a uniform rectangular array spaced at $\lambda/2$, where $\lambda=c/f_\mathrm{c}$ is the carrier wavelength. The true BS biases are equally spread in a range of $\pm 30$ m ($\pm 100$ ns), while the MT biases are spread in a small range between $\pm 0.5$ m and the true delay biases drift over time following a zig-zag pattern with a slope of $0.01$ ns. We use $10^4$ particles. 
The particles for the initial \ac{mt} position and velocity are drawn from i.i.d. Gaussian distributions with center being drawn from the same distribution for each realization, which is centered around the true \ac{mt} state, and initialization standard deviations given as $0.5~\text{m}$ and $0.1~\text{m/s}$. The initial orientation is determined from the velocity, assuming the agent to move in forward-facing direction. 
When $n \rmv=\rmv 0$, the number of \acp{va} is $0$, i.e., no prior map information is available. 
The prior for new \ac{pva} states $f_\text{n}\big(\overline{\V{x}}^{(j)}_{m,n}|\V{x}_n\big)$ is uniform on the square region given by $[-\text{45 m}, \text{45 m}] \times [-\text{45 m}, \text{45 m}]$ around the center of the floor plan shown in Fig.~\ref{fig:converging_pvas} and the mean number of new \acp{pva} at time $n$ is  $\mu_\text{n} = 0.01$. As a proposal distribution for the prior we use anulus shaped distributions, centered around the measurement for numerical efficiency. 
The probability of survival is $p_{\mathrm{s}} = 0.999$. The confirmation threshold and the pruning threshold, are given as $p_{\mathrm{cf}} = 0.5$ and $p_{\mathrm{pr}} = 10^{-3}$, respectively. 
The BS synchronization biases initialized uniformly distributed between $-100~\mathrm{ns}$ and $100~\mathrm{ns}$ 
and the MT synchronization biases are initialized uniformly distributed between $-1~\mathrm{ns}$ and $1~\mathrm{ns}$. 
The \ac{mt} state transition \ac{pdf} $f(\V{x}_{n}|\V{x}_{n-1})$ is modeled independently for MT bias, MT orientation and the MTs position-velocity state $[\RV{p}_n, \RV{v}_n]$. The MT position-velocity state is modeled by a constant-velocity and stochastic-acceleration model \cite{BarShalomBook:Book2001} with acceleration variances set to $\sigma_\text{w} = 10^{-3}\,\mathrm{m/s^2}$. The orientation state transition is modeled by a Gaussian random walk with standard deviation of  $3^\circ$. The state transitions of both, the MT biases and the BS biases, are modeled by a Gaussian random walk with driving noise set to $\sigma_{b\, BS}$ = $\sigma_{b\, MT}$ = $0.1~\text{ns}$. 
The localization and mapping performance is measured in terms of the mean \ac{rmse} of the \ac{mt} positions as well as the \acf{ospa} error \cite{Schuhmacher2008} of all \acp{va} with cutoff parameter set to $1~\mathrm{m}$ and order set to $2$. The mean \ac{ospa} (MOSPA) errors and \acp{rmse} are obtained using $100$ simulation runs.

\subsubsection*{Experiment}\label{sec:ResultsModel}

\begin{figure*}[!t]
	\centering
	\scalebox{0.8}{\includegraphics{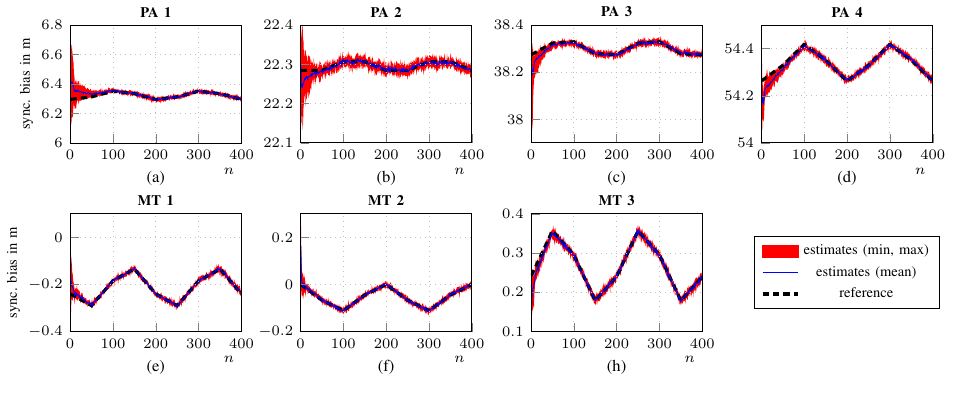}}
	\vspace*{-2mm}
	\caption{Performance of \ac{mt} and \ac{bs} biases in terms of mean error per time $n$.}\label{fig:bias}
	\vspace*{-5mm}
\end{figure*}

\begin{figure}[t]
	\centering
	\scalebox{0.9}{\includegraphics{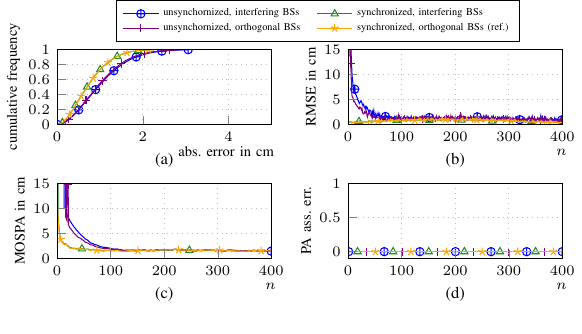}}
	\vspace*{-12mm}
	\caption{Performance in terms of (a) the cumulative frequency and (b) average \ac{rmse} of the MT position error, (c) \ac{pva} \ac{mospa}, and (d) mean number of PA association errors per time $n$.}\label{fig:results}
	\vspace*{-5mm}
\end{figure}

To investigate the proposed method, we evaluate four algorithm variants: (i) unsynchronized and interfering BSs, (ii) synchronized and interfering BSs, (iii) unsynchronized and orthogonal BSs, and (iv) synchronized and orthogonal BSs, with the (iv) being the reference case that corresponds to state-of-the-art (SOTA) cooperative \ac{mpslam} according to \cite{LeiWieVenWit:Asilomar2024}. 
Here, ``unsynchronized'' refers to unknown delay biases between BSs, while ``interfering'' indicates that the BSs use identical transmission sequences and, thus, it is unknown which MPC in the received radio signal (and, hence, which \ac{pva}) corrresponds to which BS. 
For unsychronized and interfering BS, Figure~\ref{fig:converging_pvas} shows the true PVAs and MT positions along with their respective MMSE estimates of the proposed algorithm for a single realization. One can obtain that the PVA estimates converge to the true colors, which is possible although there is no prior information about the delay biases since the \ac{aod} measurements depend on the (known) BS orientation (see Section~\ref{sec:signal_model}). Note that without \ac{aod} measurements the proposed algorithm could still differentiate, which VA belongs to which BS, but the colors would be different to the ground truth. 
The mean RMSE of \ac{bs} biases and \ac{mt} biases for $n\in\{101, \cdots, 400\}$ was $26~\mathrm{ps}$ and $24~\mathrm{ps}$, respectively. The mean error of \ac{mt} and \ac{bs} biases are illustrated in Figure~\ref{fig:bias}. Figure~\ref{fig:results} provides the numerical results for the experiment in terms of (a) the cumulative frequency and (b) average \ac{rmse} of the MT position error, (c) \ac{pva} \ac{mospa}, and (d) mean number of PA association errors per time $n$. 
One can observe that the proposed algorithm takes several times $n$ to fully converge, leading to an increased MT position error and \ac{pva} \ac{mospa} for the first $100$ steps $n$, but then reaches the performance of the reference algorithm (synchronized, orthogonal BSs). For the investigated setup, there are no \ac{pa} association errors at any time $n$ since all detected PAs correspond to the correct \ac{bs} and thus, there is no significant performance loss for unsynchronized BSs.

\vspace{-2mm} 
\section{Conclusion}\label{sec:conclusion}

\acresetall
This paper presented a Bayesian factor-graph framework for cooperative \ac{mpslam} with unsynchronized and interfering base stations. The proposed method jointly performs synchronization, source separation, and environment mapping via sequential message passing on a structured \ac{spa}. Simulation results show no significant performance degradation of the proposed algorithm compared to state-of-the-art cooperative \ac{mpslam} that assumes synchronized \acp{bs} and no inter-BS interference. 
\textit{The proposed BS-dependent data association model constitutes a principled approach for classifying features by arbitrary properties that persist over time}, such as reflection order or feature type (scatter points versus walls).
The \ac{spa} framework is flexible and can incorporate additional sensing modalities such as IMU or radar, enabling seamless integration into future 6G ISAC systems. 
Future work will address real-world experiments with 5G PRS data, extensions to dynamic network scenarios or to integrate particle flow to improve the birth of new features through the highly nonlinear models \cite{WieLeiMeyWitTSIPN2023,ZhaMey:J24}.
%

 
 \acrodef{mimo}[MIMO]{multiple input multiple output}
 \acrodef{awgn}[AWGN]{additive white Gaussian noise}
 \acrodef{bw}[BW]{bandwidth}
 \acrodef{blt}[BLT]{bluetooth}
 \acrodef{cdf}[CDF]{cumulative distribution function}
 \acrodef{crlb}[CRLB]{Cram\'er-Rao lower bound}
 \acrodef{dmc}[DMC]{dense multipath component}
 \acrodef{dut}[DUT]{device under test}
 \acrodef{eirp}[EIRP]{equivalent isotropic radiated power}
 \acrodefplural{esl}[ESLs]{electronic shelf labels} 
 \acrodef{los}[LOS]{line-of-sight}
 \acrodef{mf}[MF]{matched filter}
 \acrodef{ml}[ML]{maximum likelihood}
 \acrodef{mpc}[MPC]{multipath component}
 \acrodef{nlos}[NLOS]{non-line-of-sight}
 \acrodef{pcb}[PCB]{printed circuit board}
 \acrodef{pdf}[PDF]{probability density function}
 \acrodef{reb}[REB]{ranging error bound}
 \acrodef{rss}[RSS]{received signal strength}
 \acrodef{smc}[SMC]{specular multipath component}
 \acrodef{snr}[SNR]{signal-to-noise-ratio}
 \acrodef{sinr}[SINR]{signal-to-interference-plus-noise-ratio}
 \acrodef{tdoa}[TDOA]{time difference of arrival}
 \acrodef{tka}[TKA]{trusted keyless access}
 \acrodef{toa}[TOA]{time-of-arrival}
 \acrodef{aoa}[AOA]{angle-of-arrival}
 \acrodef{uwb}[UWB]{ultra wide band}
 \acrodef{mie}[MIE]{method of interval estimation}
 \acrodef{mc}[MC]{Monte Carlo}
 \acrodef{mse}[MSE]{mean squared error}
 \acrodef{ci}[CI]{confidence interval}
 \acrodef{cl}[CL]{confidence level}
 \acrodef{pdp}[PDP]{power delay profile}
 \acrodef{dps}[DPS]{delay power spectrum}
 \acrodef{dm}[DM]{dense multipath}
 \acrodef{nlike}[NLIKE]{normalized likelihood}
 \acrodef{zzb}[ZZB]{Ziv-Zakai bound}
 \acrodef{ut}[UT]{unscented transform}
 \acrodef{glrt}[GLRT]{generalized likelihood ratio test}
 \acrodef{mse}[MSE]{mean squared error}
 \acrodef{rmse}[RMSE]{root mean squared error}
 \acrodef{nnlike}[NNLIKE]{normalized noise-free likelihood}
 \acrodef{stdv}[STDV]{standard deviation}
 \acrodef{rv}[RV]{random variable}
 \acrodef{bp}[BP]{belief propagation}
 \acrodef{pda}[PDA]{probabilistic data association}
 \acrodef{mp}[MP]{multipath}
 \acrodef{pmf}[PMF]{probability mass function}
 \acrodef{pdaf}[PDAF]{probabilistic data association filter}
 \acrodef{pdaai}[AIPDA]{amplitude-information \ac{pda}}
 \acrodef{olos}[OLOS]{obstructed line-of-sight}
 \acrodef{spa}[SPA]{sum-product algorithm}
 \acrodef{mmse}[MMSE]{minimum mean-square error}
 \acrodef{lhf}[LHF]{likelihood function}
 \acrodef{fa}[FA]{false alarm}
 \acrodef{fp}[FP]{false positive}
 \acrodef{ceda}[CEDA]{channel estimation and detection algorithm} 
 \acrodef{pcrlb}[P-CRLB]{posterior Cram\'er-Rao lower bound}
 \acrodef{mpslam}[MP-SLAM]{multipath-based SLAM}
 \acrodef{va}[VA]{virtual anchor}
  \acrodef{pva}[PVA]{potential virtual anchor}
 \acrodef{dnr}[DNR]{dense-to-noise ratio}
 \acrodef{pbo}[PBO]{potential bias object}
 \acrodef{npbo}[NPBO]{new \ac{pbo}}
 \acrodef{lpbo}[LPBO]{legacy \ac{pbo}}
 \acrodef{aednn}[AE-DNN]{autoencoder deep neural network}   
 \acrodef{gpr}[GPR]{Gaussian process regression}  
 \acrodef{cluster}[CLUSTER]{{\color{red}error}}  
 \acrodef{delaybias}[ML-BIAS]{{\color{red}error}}  
 \acrodef{gptrack}[GP-TRACK]{{\color{red}error}}  
 \acrodef{chslam}[CH-SLAM]{{\color{red}error}}  
 \acrodef{wrt}[w.r.t.]{with respect to} 
 \acrodef{pa}[PA]{physical anchor} 
\acrodef{bs}[BS]{base station}
\acrodef{mt}[MT]{mobile terminal}
\acrodef{aoa}[AOA]{angle-of-arrival}
\acrodef{aod}[AOD]{angle-of-departure}
\acrodef{mpslam}[MP-SLAM]{multipath-based simultaneous localization and mapping}
\acrodef{slam}[SLAM]{simultaneous localization and mapping}
\acrodef{rf}[RF]{radio frequency}
\acrodef{imu}[IMU]{inertial measurement unit}
\acrodef{ospa}[OSPA]{optimal subpattern assignment}
\acrodef{mospa}[MOSPA]{mean optimal subpattern assignment}


\renewcommand{\baselinestretch}{0.93}\small\normalsize

\vspace{-2mm} 
\bibliographystyle{IEEEbib}
\bibliography{IEEEabrv,References,TempRefs}

\end{document}